\documentclass[12pt, draftclsnofoot, journal, onecolumn]{IEEEtran}
\makeatletter

\usepackage{blindtext}
\usepackage[T1]{fontenc}
\usepackage{textcomp}
\usepackage{amsmath}
\usepackage{amssymb}
\usepackage{bm}
\usepackage{mdwmath}
\usepackage{cases}
\usepackage{graphicx}
\graphicspath{{Figure/PDF/}{Biography/PDF/}}
\usepackage[dvipsnames]{xcolor}
\definecolor{myblue}{rgb}{0,0.4980,1} 
\definecolor{myred}{rgb}{0.8706,0.1608,0.0627} 
\usepackage[caption=false,font=footnotesize,subrefformat=parens]{subfig}
\usepackage[nosort,noadjust]{cite}
\usepackage{url}
\usepackage{tabularray}
\UseTblrLibrary{counter}

\usepackage{hyperref}
\newcommand{\colorhypersetup}{\@ifpackageloaded{hyperref}{\hypersetup{%
	bookmarksopen=true,%
	bookmarksnumbered=true,%
	pdfpagemode={UseOutlines},
	pdfstartview={FitH},%
	colorlinks=true,%
	linkcolor={myred},%
	citecolor={orange}
}}{\empty}}
\newcommand{\blackhypersetup}{\@ifpackageloaded{hyperref}{\hypersetup{%
	bookmarksopen=true,%
	bookmarksnumbered=true,%
	pdfpagemode={UseOutlines},
	pdfstartview={FitH},%
	colorlinks=true,%
	allcolors={black}
}}{\empty}}
\colorhypersetup

\usepackage{mfirstuc-english}
\MFUhyphentrue
\usepackage[version2,description,IEEEtran]{eacro}
\acsetup{%
    page-style=paren,%
    hyperref=false%
}
\DeclareAcronym{ACB}{
	short = ACB,
	long = access class barring}
\DeclareAcronym{CDF}{
	short = CDF,
	long = cumulative distribution function}
\DeclareAcronym{DDQN}{
	short = DDQN,
	long = double deep Q network}
\DeclareAcronym{EMA}{
	short = EMA,
	long = exponential moving average}
\DeclareAcronym{FIFO}{
	short = FIFO,
	long = first-in first-out}
\DeclareAcronym{HPBW}{
	short = HPBW,
	long = half-power beamwidth}
\DeclareAcronym{MDP}{
	short = MDP,
	long = Markov decision process}
\DeclareAcronym{RAR}{
	short = RAR,
	long = random access response}
\DeclareAcronym{ReLU}{
	short = ReLU,
	long = rectified linear unit}
\DeclareAcronym{RRC}{
	short = RRC,
	long = radio resource control}
\DeclareAcronym{ULA}{
	short = ULA,
	long = uniform linear array}
\DeclareAcronym{PRACH}{
	short = PRACH,
	long = physical random access channel}
\DeclareAcronym{M2M}{
	short = M2M,
	long = machine-to-machine}
\DeclareAcronym{DDPG}{
	short = DDPG,
	long = deep deterministic policy gradient}
\DeclareAcronym{IoT}{
	short = IoT,
	long = Internet of things}
\DeclareAcronym{MTC}{
	short = MTC,
	long = machine-type communication}
\DeclareAcronym{3GPP}{
	short = 3GPP,
	long = 3rd generation partnership project}
\DeclareAcronym{umMTC}{
	short = umMTC,
	long = ultra-massive machine-type communication}
\DeclareAcronym{5G}{
	short = 5G,
	long = 5th generation}
\DeclareAcronym{6G}{
	short = 6G,
	long = 6th generation}
\DeclareAcronym{H2H}{
	short = H2H,
	long = human-to-human}
\DeclareAcronym{RACH}{
	short = RACH,
	long = random access channel}
\DeclareAcronym{MIMO}{
	short = MIMO,
	long = multi-input multi-output}
\DeclareAcronym{CSI}{
	short = CSI,
	long = channel state information}
\DeclareAcronym{LTE}{
	short = LTE,
	long = Long-Term Evolution}
\DeclareAcronym{MSE}{
	short = MSE,
	long = mean squared error}

\usepackage[ruled,linesnumbered,commentsnumbered]{algorithm2e}
\SetAlCapHSkip{0pt}
\SetAlgoCaptionSeparator{.}
\SetKwInput{Kwflow}{Procedure}
\usepackage[per-mode=symbol,group-separator={,}]{siunitx}

\newtheorem{remark}{\textbf{Remark}}

\newcommand{\upperroman}[1]{\uppercase\expandafter{\romannumeral#1}}

\newcommand{\myvec}[1]{\bm{\mathrm{#1}}}

\DeclareMathOperator{\argmax}{argmax}
\newcommand{\myunit}[1]{%
	\ifmmode
		\mathrm{#1}
	\else
		$ \mathrm{#1} $
	\fi}
\newcommand{\murm}{%
	\ifmmode
		\text{\textmu}
	\else
		\textmu
	\fi}

\newcommand{\MYnewpage}{%
	\ifCLASSOPTIONonecolumn
		\ifCLASSOPTIONjournal
			\typeout{The onecolumn journal mode.}
			\newpage
		\fi
	\fi}

\newlength{\mysinglefigwidth}
\newlength{\mymultifigwidth}
\ifCLASSOPTIONonecolumn
	\AtBeginDocument{\setlength{\mysinglefigwidth}{0.7\linewidth}}
\else
	\AtBeginDocument{\setlength{\mysinglefigwidth}{\linewidth}}
\fi
\ifCLASSOPTIONonecolumn
	\AtBeginDocument{\setlength{\mymultifigwidth}{0.5\linewidth}}
\else
	\AtBeginDocument{\setlength{\mymultifigwidth}{\linewidth}}
\fi

\makeatother
\begin{document}
\ifCLASSOPTIONonecolumn
	\typeout{The onecolumn mode.}
	\title{Dynamic Beam-Based Random Access Scheme for M2M Communications in Massive MIMO Systems}
	\author{Kan~Zheng,~\IEEEmembership{Senior~Member,~IEEE}, Haojun~Yang,~\IEEEmembership{Member,~IEEE}, Xiong~Xiong, Jie~Mei~\IEEEmembership{Member,~IEEE}, and~Kuan~Zhang,~\IEEEmembership{Member,~IEEE}
	}
\else
	\typeout{The twocolumn mode.}
	\title{Dynamic Beam-Based Random Access Scheme for M2M Communications in Massive MIMO Systems}
	\author{Kan~Zheng,~\IEEEmembership{Senior~Member,~IEEE}, Haojun~Yang,~\IEEEmembership{Member,~IEEE}, Xiong~Xiong, Jie~Mei~\IEEEmembership{Member,~IEEE}, and~Kuan~Zhang,~\IEEEmembership{Member,~IEEE}
	}
\fi

\ifCLASSOPTIONonecolumn
	\typeout{The onecolumn mode.}
\else
	\typeout{The twocolumn mode.}
	\markboth{IEEE Transactions on Vehicular Technology}{Author \MakeLowercase{\textit{et al.}}: Title}
\fi

\maketitle

\ifCLASSOPTIONonecolumn
	\typeout{The onecolumn mode.}
	\vspace*{-50pt}
\else
	\typeout{The twocolumn mode.}
\fi
\begin{abstract}
Internet of things, supported by machine-to-machine (M2M) communications, is one of the most important applications for future 6th generation (6G) systems. A major challenge facing by 6G is enabling a massive number of M2M devices to access networks in a timely manner. Therefore, this paper exploits the spatial selectivity of massive multi-input multi-output (MIMO) to reduce the collision issue when massive M2M devices initiate random access simultaneously. In particular, a beam-based random access protocol is first proposed to make efficient use of the limited uplink resources for massive M2M devices. To address the non-uniform distribution of M2M devices in the space and time dimensions, an Markov decision process (MDP) problem with the objective of minimizing the average access delay is then formulated. Next, we present a dynamic beam-based access scheme based on the double deep Q network (DDQN) algorithm to solve the optimal policy. Finally, simulations are conducted to demonstrate the effectiveness of the proposed scheme including the model training and random access performance.
\end{abstract}

\ifCLASSOPTIONonecolumn
	\typeout{The onecolumn mode.}
	\vspace*{-10pt}
\else
	\typeout{The twocolumn mode.}
\fi

\IEEEpeerreviewmaketitle

\MYnewpage


\section{Introduction}
\label{sec:introduction}

\acresetall

\IEEEPARstart{R}{ecently}, various types of applications in \ac{IoT} have rapidly grown, especially for industrial field. A high number of end devices are deployed, and adopt \ac{IoT} networks to transmit sensing data and control information in a timely manner. To well support \ac{IoT} applications, the \acs{3GPP} has developed \acp{MTC} and \acp{umMTC} as typical use cases in \ac{5G} and \ac{6G} networks~\cite{Varsier2021,Tataria2022}. Consequently, the cellular-enabled \ac{IoT} networks are receiving considerable attention from the industry and academia for operating diverse \ac{IoT} applications.

The current mobile cellular networks, however, may not be able to meet the requirements of either \acp{MTC} or \acp{umMTC}, because they were originally designed to support human-type communications. In contrast to \ac{H2H} communications, \ac{M2M} communications typically involve a massive number of devices with low-data rate transmitting small payloads sporadically~\cite{Tanab2021}. Moreover, \ac{MTC} devices usually are not uniformly deployed across geographical areas, due to various \ac{IoT} applications existing in networks. Based on the unique features and location distribution of \ac{MTC} traffic, the current cellular networks need to be greatly improved to efficiently handle \ac{M2M} communications. In general, \ac{M2M} devices are first required to establish air interface connections in cellular networks before data transmission. The corresponding access requests are transmitted in an uncoordinated manner over the \acp{RACH}. Once an \ac{MTC} device has been granted to access, it is scheduled to use specific radio resources over which data transmission takes place in a deterministic manner. However, when a massive number of \ac{MTC} devices are deployed in the networks, the \acp{RACH} with the existing access schemes become seriously overloaded result in heavy congestion. To this end, one of key issues is how to guarantee the timely access requirements of massive \ac{M2M} devices.

Some recent advances have been conducted on dealing with the previous challenge~\cite{laya_is_2014, ghavimi_m2m_2015}. First of all, the number of \ac{M2M} devices that initiate random access at the same time can be limited by a pre-defined access probability in \ac{ACB}, thereby reducing the risk of collision~\cite{3GPP_tr37.868_2011, lien_cooperative_2012, wang_optimal_2015, duan_d-acb_2016, jin_recursive_2017}. Then, by delaying the access time of the \ac{M2M} devices with delay-insensitive applications, all random access requests can be scheduled into different time slots, which can relieve \ac{PRACH} congestion~\cite{yang_performance_2012, chen_dynamic_2020, althumali_dynamic_2020}. Meanwhile, the base station can dynamically allocate radio resources to \ac{M2M} devices based on the conditions of \acp{PRACH} and network loads~\cite{li_dynamic_2015, hwang_dynamic_2015}. Additionally, the base station can fully control random access by adjusting the paging timing of \ac{M2M} devices ~\cite{ghavimi_m2m_2015}. A dedicated time slot of random access also can be assigned to reduce the interference on \ac{H2H} communications~\cite{3GPP_tr37.868_2011, shahin_hybrid_2018}. Finally, the code-expanded access is another good method~\cite{pratas_code-expanded_2012}. For instance, in order to reduce preamble collision, a virtual preamble can be created by integrating the classical preamble and \ac{PRACH} channel index~\cite{kim_efficient_2017}. Although the previous methods can help to reduce the congestion in the cellular mobile network, the issue concerning the massive number of \ac{M2M} devices is still not solved effectively.

In recent years, massive \ac{MIMO} has become one of the key techniques in \ac{5G} and \ac{6G} networks. The challenges of massive \ac{M2M} devices accessing massive \ac{MIMO} systems have been investigated. For example, an approximate message passing based grant-free scheme is proposed to jointly detect active users and estimate \ac{CSI}~\cite{Liu2018}. Furthermore, a grant-free random access scheme is studied to detect active \ac{M2M} devices and uplink messages without \ac{CSI} estimation in one shot~\cite{Han2020}. With the aid of the spatial multiplexing concept, multiple users and devices can use the same time-frequency resources simultaneously, thereby improving system throughput and spectral efficiency~\cite{rusek_scaling_2013, larsson_massive_2014}. However, few schemes have explored how to use the spatial characteristics of massive \ac{MIMO} to solve the random access problem of massive \ac{M2M} devices~\cite{xiong_group-based_2018}. Intuitively, the base station with a large antenna array can generate a huge number of beams in different directions. Due to the high spatial resolution of beams, mutually isolated spatial areas in specific directions is feasible for the cellular network, which is more beneficial to let a certain beam only serve the devices within its corresponding area~\cite{carvalho_random_2017}.  

To address the access problem of massive \ac{M2M} devices located at different geographic locations, this paper proposes a beam-based dynamic random access scheme based on massive \ac{MIMO} technique. In particular, some \ac{M2M} devices located in the coverage of a certain beam can be divided into the same group, and can use the same set of preambles and radio resources. To further improve the utilization efficiency, the same set of preambles and radio resources can also be reused by the \ac{M2M} devices in the different beams. Owing to the spatial orthogonality of different beams, the mutual interference among the \ac{M2M} devices in different groups is extremely small. Therefore, the collision problem can be effectively alleviated, when a massive number of \ac{M2M} devices initiate the access simultaneously

The main contributions of this paper include:
\begin{itemize}
\item We propose a beam-based random access protocol for massive \ac{M2M} devices. A group of beams with unequal beamwidth are applied to address the impact of the non-uniform distribution of \ac{M2M} devices in space and time on random access process.

\item Based on the \ac{MDP}, an optimization problem with the objective of minimizing the average access delay is formulated. To solve the optimal policy, we develop a dynamic beam-based random access scheme based on the \ac{DDQN} algorithm.

\item In order to demonstrate the effectiveness of the proposed scheme, the model training performance and random access performance are compared with other reference random access schemes by the simulation results.
\end{itemize}

The rest of this paper is organized as follows. In Section~\ref{sec:model}, a typical scenario and model are first presented. Then, a beam-based random access protocol for \ac{M2M} devices is described in detail. Next, a dynamic beam-based random access problem is formulated as \ac{MDP} model in Section~\ref{sec:MDP}. Based on the \ac{DDPG}, Section~\ref{sec:solution} proposes an algorithm to find the optimal policy. Moreover, simulations are carried out to demonstrate the performance of the proposed dynamic scheme in Section~\ref{sec:simulation}. Finally, the conclusions are presented in Section~\ref{sec:conclusion}.

\section{System Model and Random Access Protocol}
\label{sec:model}

\subsection{Scenario Description}
\label{subsec:scenario}

\begin{figure}[!t]
	\centering
	\includegraphics[width=\mysinglefigwidth]{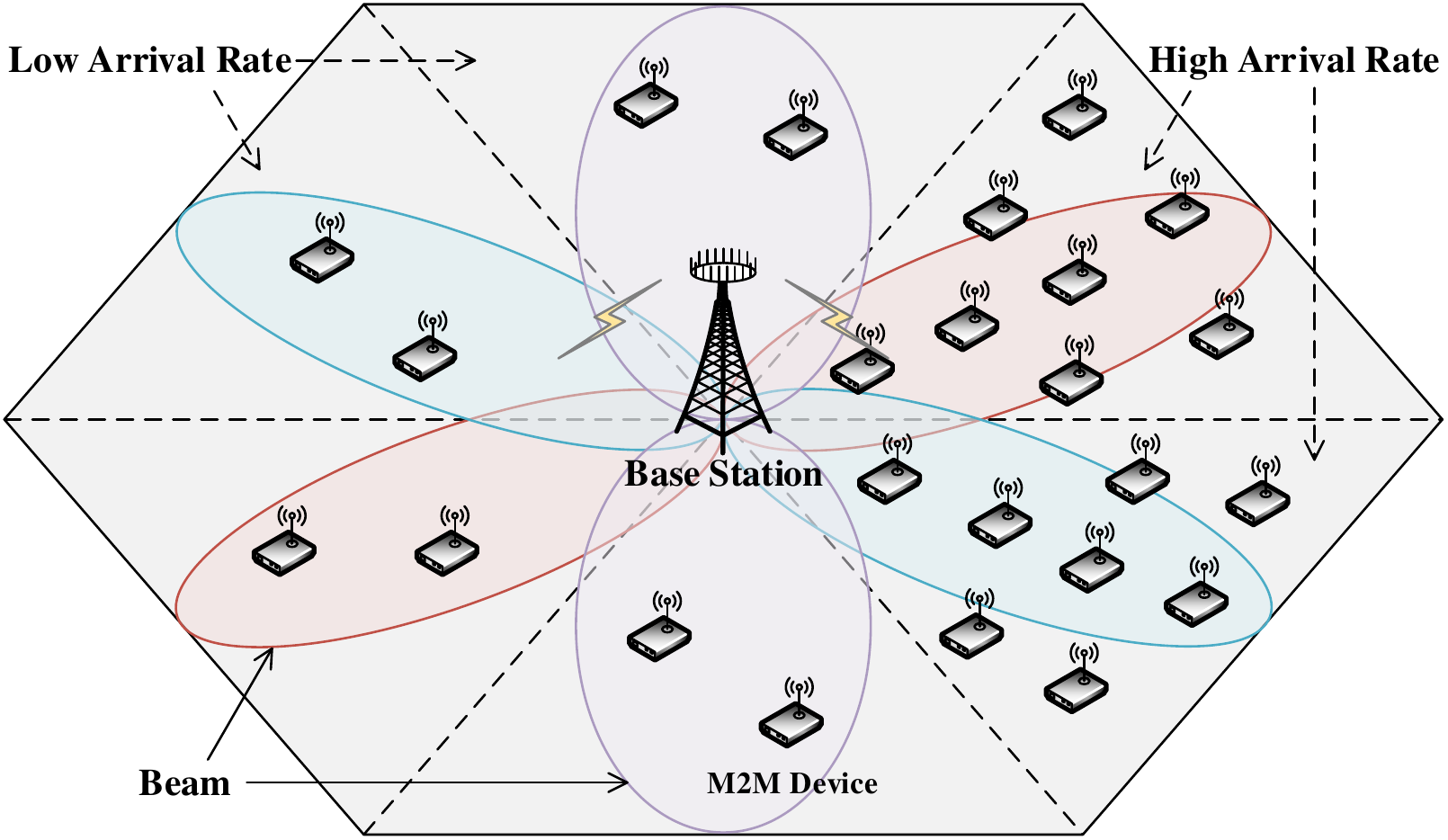}
	\caption{Illustration of a base station with massive \acs*{MIMO} serving \acs*{M2M} devices.}
	\label{fig:scenario}
\end{figure}

As shown in Fig.~\ref{fig:scenario}, $N_{\mathrm{g}}$ beams generated by a base station with massive \ac{MIMO} are adopted to communicate with \ac{M2M} devices located in the different areas. Furthermore, a set of $N_{\mathrm{p}}$ preambles are used for random access within a single beam, but might be reused between different beams. In general, the spatial distribution of \ac{M2M} devices is usually not uniform and is also highly dynamic. This results in different arrival rates of random access both in space dimension and time dimension. For purposes of analysis, the coverage area of a base station is divided into $N_{\mathrm{s}}$ equal sectors. The \ac{M2M} devices are uniformly distributed in a single sector, and their random access requests follow a Poisson distribution, where the arrival rates of access can be denoted as $\lambda_{j}, {j \in \{1, 2, \dotsc, N_{\mathrm{s}}\}}$.

\subsection{Description of Beam-Based Random Access Protocol}
\label{subsec:group_ra}

In this paper, a beam-based random access protocol for massive \ac{M2M} devices is proposed, and it mainly consists of the following four steps.

\subsubsection{\textbf{Step 1 -- Preamble Transmission}}
\label{subsubsec:step1}
Similarly with the regular random access protocol in \ac{LTE} systems, in our proposed protocol, a \ac{M2M} device randomly selects and sends a preamble to the base station on the first available time slot of random access.

\subsubsection{\textbf{Step 2 -- Random Access Response}}
\label{subsubsec:step2}

After detecting the preamble sent by the \ac{M2M} device, the base station sends the corresponding \ac{RAR} in the downlink, including the index of the detected preamble, the indicator of uplink resource and other information. When more than one \ac{M2M} device send the same preamble, the base station will send the same \ac{RAR} to them. However, it cannot tell whether a collision happens now, which has to be addressed in the subsequent steps.

\subsubsection{\textbf{Step 3 -- Connection Request}}
\label{subsubsec:step3}

Based on the received \ac{RAR}, the \ac{M2M} device sends an \ac{RRC} message to the base station. When the base station is equipped with massive \ac{MIMO}, it can distinguish various \ac{RRC} messages from the \ac{M2M} devices located at different areas by multiple directional beams. Specifically, since most of the useful signals radiated by the antennas are concentrated within a beam, the beam has the remarkable spatial selectivity. Thus, the \ac{M2M} devices located in the different beam coverage can be served separately, thereby reducing mutual interference. That is to say, for those \ac{M2M} devices that send the same preamble and receive the same \ac{RAR}, the base station can also distinguish the \ac{RRC} messages through exploiting the spatial selectivity of different beams. Therefore, the beam-based random access scheme can efficiently improve the successful probability of random access.

\subsubsection{\textbf{Step 4 -- Contention Resolution}}
\label{subsubsec:step4}

After the base station successfully demodulates the \ac{RRC} messages, it sends a contention resolution message back to the corresponding \ac{M2M} device. Only the \ac{M2M} device whose \ac{RRC} message has been successfully demodulated can confirm the contention resolution message message and complete the random access procedure, while other \ac{M2M} devices fail to access.

\subsection{Formulation of Beam-Based Random Access Protocol}

\begin{figure}[!t]
	\centering
	\includegraphics[width=0.8\mysinglefigwidth]{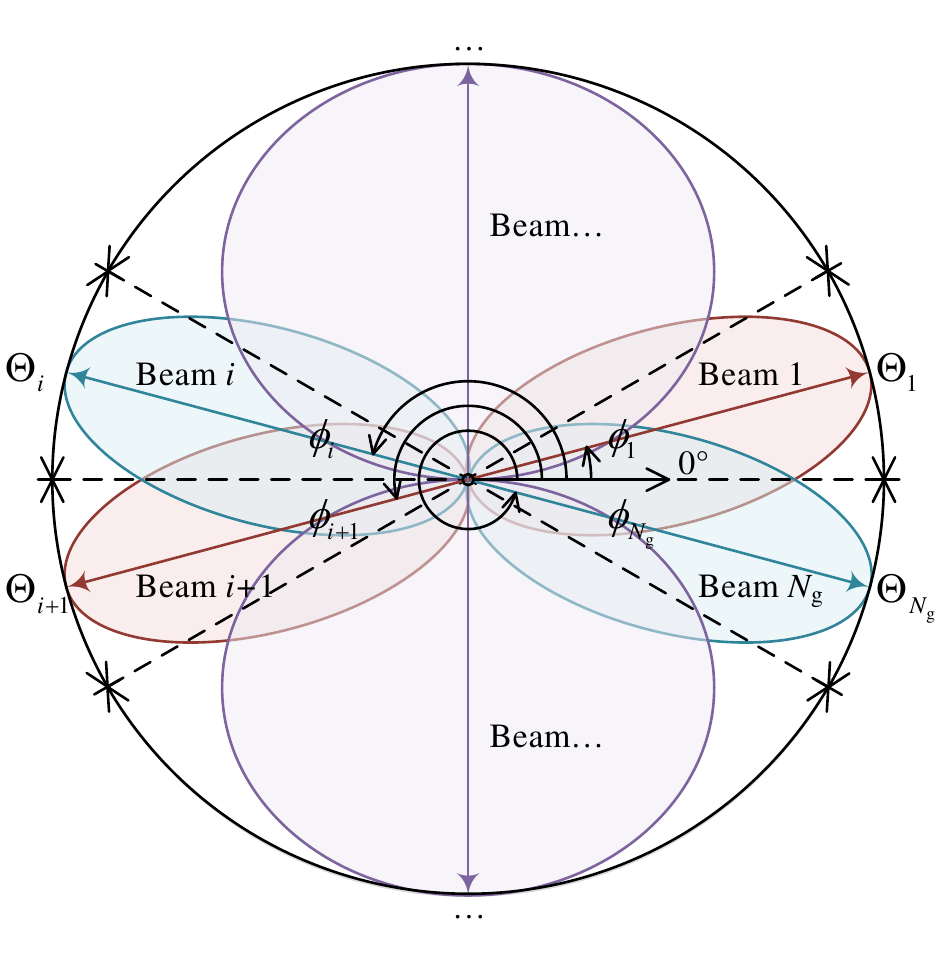}
	\caption{Illustration of beam pattern with massive \acs*{MIMO}.}
	\label{fig:beams}
\end{figure}

\begin{algorithm}[!t]
\setlength{\algowidth}{\hsize}
\setlength{\hsize}{\linewidth}
\caption{Dynamic beam-based random access protocol.}
\label{alg:beam}
\setlength{\hsize}{\algowidth}
\setlength{\algowidth}{\linewidth}
\KwIn{}
The number of beams $N_{\mathrm{g}}$\;
The number of preambles $N_{\mathrm{p}}$\;
The arrival rates of access for $ N_{\mathrm{s}} $ sectors, namely $\lambda_{j}$, ${j \in \{1, 2, \dotsc, N_{\mathrm{s} }\}}$.
\BlankLine
\Kwflow{}
\textbf{Step 1 -- Preamble Transmission:} The \ac{M2M} devices randomly select and send preambles to the base station\;
\textbf{Step 2 -- Random Access Response:} The base station first performs dynamic random access scheme as described in Algorithm~\ref{alg:model.DDQN}, and then send the corresponding \ac{RAR}\;
\textbf{Step 3 -- Connection Request:} The \ac{M2M} devices send the corresponding \ac{RRC} messages to the base station\;
\textbf{Step 4 -- Contention Resolution:} When the \ac{RRC} messages are successfully decoded by the base station, i.e., the requirement in \eqref{eq: msg3_decoding} is met, the \ac{M2M} devices can receive the contention resolution message, and access to the network successfully.
\BlankLine
\If{The \ac{M2M} devices fail to access}{\textbf{Repeat} Step~1--4\;}
\end{algorithm}

In this section, we elaborate the proposed protocol mathematically. As shown in Fig.~\ref{fig:beams}, each beam can be characterized by the maximum gain direction of $\phi_{i}$ and beamwidth of $\Theta_{i}$. Generally, a beam is symmetric about its maximum gain direction, and most of its gain is concentrated within the beamwidth. In order to fully cover the service area of the base station, these beams need to be adjacent to each other with their beamwidth. In the other words, the maximum gain direction and beamwidth of each beam should meet the following constraints, i.e.,
\begin{subnumcases}{\label{eq:beam_constraint}}
\text{\textbf{C1:}}\quad \sum\limits_{i=1}^{N_{\mathrm{g}}} \Theta_{i} = 2\pi, \\
\text{\textbf{C2:}}\quad \phi_{(i \bmod{N_{\mathrm{g}}}) +1} = \Big[\phi_{i} + \notag \\
\phantom{\phi_{(i \bmod{N_{\mathrm{g}}})}} \left(\Theta_{(i \bmod{N_{\mathrm{g}}}) + 1} + \Theta_{i} \right)/2 \Big] \bmod{2\pi}, \\
\text{\textbf{C3:}}\quad 0 \leq \phi_i, \Theta_i  < 2\pi , i \in \{1, 2, \dotsc, N_{\mathrm{g}}\},
\end{subnumcases}
where $\bmod (\cdot)$ represents the modulo operation.

Without loss of generality, let $(\theta_0, d_0)$ be the location of the interested \ac{M2M} device in polar coordinate, where $\theta_0$ and $d_0$ represent the angle and distance between the device and the base station, respectively. It is regarded as having successfully connected to the network only after the base station demodulated the \ac{RRC} message through the beam $i^*$. The serving beam $i^*$ here refers to the one that the polar coordinate $(\theta_0,d_0)$ of the interested \ac{M2M} device is converged within its bandwidth range, i.e.,
\begin{subnumcases}{}
\theta_{i^*}^{\mathrm{L}} \le \theta_0 < 2\pi \lor 0 \le \theta_0 \le \theta_{i^*}^{\mathrm{H}}, & $\theta_{i^*}^{\mathrm{L}} > \theta_{i^*}^{\mathrm{H}},$ \\
\theta_{i^*}^{\mathrm{L}} \le \theta_0 \le \theta_{i^*}^{\mathrm{H}}, & $\theta_{i^*}^{\mathrm{L}} \le \theta_{i^*}^{\mathrm{H}},$
\end{subnumcases}
where $\theta_{i^*}^{\mathrm{L}}$ and $\theta_{i^*}^{\mathrm{H}}$ are lower and upper boundaries of the serving beam $i^*$. Here we take the periodicity of the angle into account, i.e., $\theta_{i^*}^{\mathrm{L}} = (\phi_{i^*} - \Theta_{i^*}/2) \bmod{2\pi}$, $\theta_{i^*}^{\mathrm{H}} = (\phi_{i^*} + \Theta_{i^*}/2) \bmod{2\pi}$.

For the interested \ac{M2M} device, the signals with \ac{RRC} messages sent by the other $N_{\mathrm{I}}$ \ac{M2M} devices served by those beams different from the beam $i^*$ are regarded as interference, because the same uplink radio resource is used. The received power of the interference at the serving beam $i^*$ for the interested device can be expressed as (in \myunit{dBm})
\begin{align}
I_{i^*} = 10 \operatorname{log}_{10}
    \left[\sum_{q=1}^{N_{\mathrm{I}}} 10^{R_{i^*}(\theta_q, d_q)/10} \right], q \in \{1, 2, \dots, N_{\mathrm{I}} \}.
\label{eq:msg3_interference_power}
\end{align}
Accordingly, at the $i$-th beam of the base station, the received power of the signal with the \ac{RRC} message are generally calculated by (in \myunit{dBm})
\begin{align}
R_i(\theta, d) &= P_{\mathrm{t}} + G_{\mathrm{t}} + G_{\mathrm{r}} + 20 \log_{10}(f_i(\theta)) \notag \\
&\phantom{P_{\mathrm{t}} + G_{\mathrm{t}} +} - \operatorname{PL}(d) + \chi, i \in \{1, 2, \dots, N_{\mathrm{g}} \},
\label{eq:msg3_signal_power}
\end{align}
where $P_{\mathrm{t}}$ represents the transmit power of the \ac{M2M} device, while $G_{\mathrm{t}}$ and $G_{\mathrm{r}}$ are the gains of transmit and receive antennas, respectively. $\operatorname{PL}(d)$ is the path loss between the \ac{M2M} device and base station, and $\chi$ is the shadow fading obeying a Gaussian distribution with zero mean and variance $\sigma^2$. Lastly, $f_i(\theta)$ represents the gain of the $i$-th beam in the direction $\theta$, and Appendix~\ref{sec:appendix} gives the calculation of the gain.

Finally, when the power difference in logarithmic value between the \ac{RRC} message and the interference is larger than the required demodulation threshold $\Gamma$, the base station can successfully decode the \ac{RRC} message of the interested \ac{M2M} device, i.e.,
\begin{align}
R_{i^*}(\theta_0, d_0) - I_{i^*} > \Gamma.
\label{eq: msg3_decoding}
\end{align}

To sum up, the dynamic beam-based random access protocol is proposed in Algorithm~\ref{alg:beam}. In the next section, we mainly focus on the study of the dynamic random access part.

\section{Problem Formulation of Dynamic Beam-Based Random Access}
\label{sec:MDP}

According to the proposed beam-based random access process, the base station can \textbf{\textit{dynamically}} adjust the direction and beamwidth of the beam based on the non-uniform distribution of \ac{M2M} devices. Thus, the number of devices located in the coverage of different beams can be effectively balanced, thereby alleviating the collision in random access process. In this section, the \textbf{\textit{dynamic}} random access problem is formulated as a \ac{MDP} model, and its state, action, and revenue are defined as follows.

\subsection{State}
\label{subsubsec:model.state}

The state $s \in \mathcal{S}$ includes the number of \ac{M2M} devices that initiate random access in each sector area, i.e.,
\begin{align}
\mathcal{S}=\left\{s \mid s = (n_{1}, n_{2}, \dotsc, n_{N_{\mathrm{s}}}) \right\},
\label{eq:model.state}
\end{align}
where $n_{j},j \in \{1, 2, \dotsc, N_{\mathrm{s}}\}$ represents the number of \ac{M2M} devices accessing in the $j$-th sector.

\subsection{Action}
\label{subsubsec:model.action}

\begin{table*}[!t]
\centering
\caption{Example of Action Space}
\label{tab:model.action}
\begin{tblr}{
    width = 0.8\linewidth,
    colspec = {X[2,c,m]X[1,c,m]X[1,c,m]X[1,c,m]X[1,c,m]X[1,c,m]X[1,c,m]X[1,c,m]X[1,c,m]X[1,c,m]X[1,c,m]X[1,c,m]X[1,c,m]},
    hlines,
    hline{3} = {1}{-}{},
    hline{3} = {2}{-}{},
    vline{2-13},
    row{1} = {font=\bfseries},
    cell{1}{1} = {r=2}{c,m},
    cell{1}{2} = {c=2}{c,m},
    cell{1}{4} = {c=2}{c,m},
    cell{1}{6} = {c=2}{c,m},
    cell{1}{8} = {c=2}{c,m},
    cell{1}{10} = {c=2}{c,m},
    cell{1}{12} = {c=2}{c,m},
}
    Action & Beam 1 & temp & Beam 2 & temp & Beam 3 & temp & Beam 4 & temp & Beam 5 & temp & Beam 6 & temp \\
    temp & $\phi_{1}$ & $\Theta_{1}$ & $\phi_{2}$ & $\Theta_{2}$ & $\phi_{3}$ & $\Theta_{3}$ & $\phi_{4}$ & $\Theta_{4}$ & $\phi_{5}$ & $\Theta_{5}$ & $\phi_{6}$ & $\Theta_{6}$ \\
    $a_{1}$ & $\dfrac{1}{12}\pi$ & $\dfrac{1}{6}\pi$ & $\dfrac{1}{2}\pi$ & $\dfrac{2}{3}\pi$ & $\dfrac{11}{12}\pi$ & $\dfrac{1}{6}\pi$ & $\dfrac{13}{12}\pi$ & $\dfrac{1}{6}\pi$ & $\dfrac{3}{2}\pi$ & $\dfrac{2}{3}\pi$ & $\dfrac{23}{12}\pi$ & $\dfrac{1}{6}\pi$ \\
    $a_{2}$ & $\dfrac{5}{12}\pi$ & $\dfrac{1}{6}\pi$ & $\dfrac{5}{6}\pi$ & $\dfrac{2}{3}\pi$ & $\dfrac{5}{4}\pi$ & $\dfrac{1}{6}\pi$ & $\dfrac{17}{12}\pi$ & $\dfrac{1}{2}\pi$ & $\dfrac{11}{6}\pi$ & $\dfrac{2}{3}\pi$ & $\dfrac{1}{4}\pi$ & $\dfrac{1}{6}\pi$ \\
    $a_{3}$ & $\dfrac{3}{4}\pi$ & $\dfrac{1}{6}\pi$ & $\dfrac{7}{6}\pi$ & $\dfrac{2}{3}\pi$ & $\dfrac{19}{12}\pi$ & $\dfrac{1}{6}\pi$ & $\dfrac{7}{4}\pi$ & $\dfrac{1}{2}\pi$ & $\dfrac{13}{6}\pi$ & $\dfrac{2}{3}\pi$ & $\dfrac{7}{12}\pi$ & $\dfrac{1}{6}\pi$
\end{tblr}
\end{table*}

\begin{figure*}[!t]
\centering
\begin{tblr}{
    width = \linewidth,
    colspec = {X[1,c,m]X[1,c,m]},
    columns = {leftsep=0pt,rightsep=0pt},
    cell{1}{1} = {c=2}{c,m},
}
	\subfloat[Action $a_{1}$.]{\includegraphics[width=0.8\mymultifigwidth]{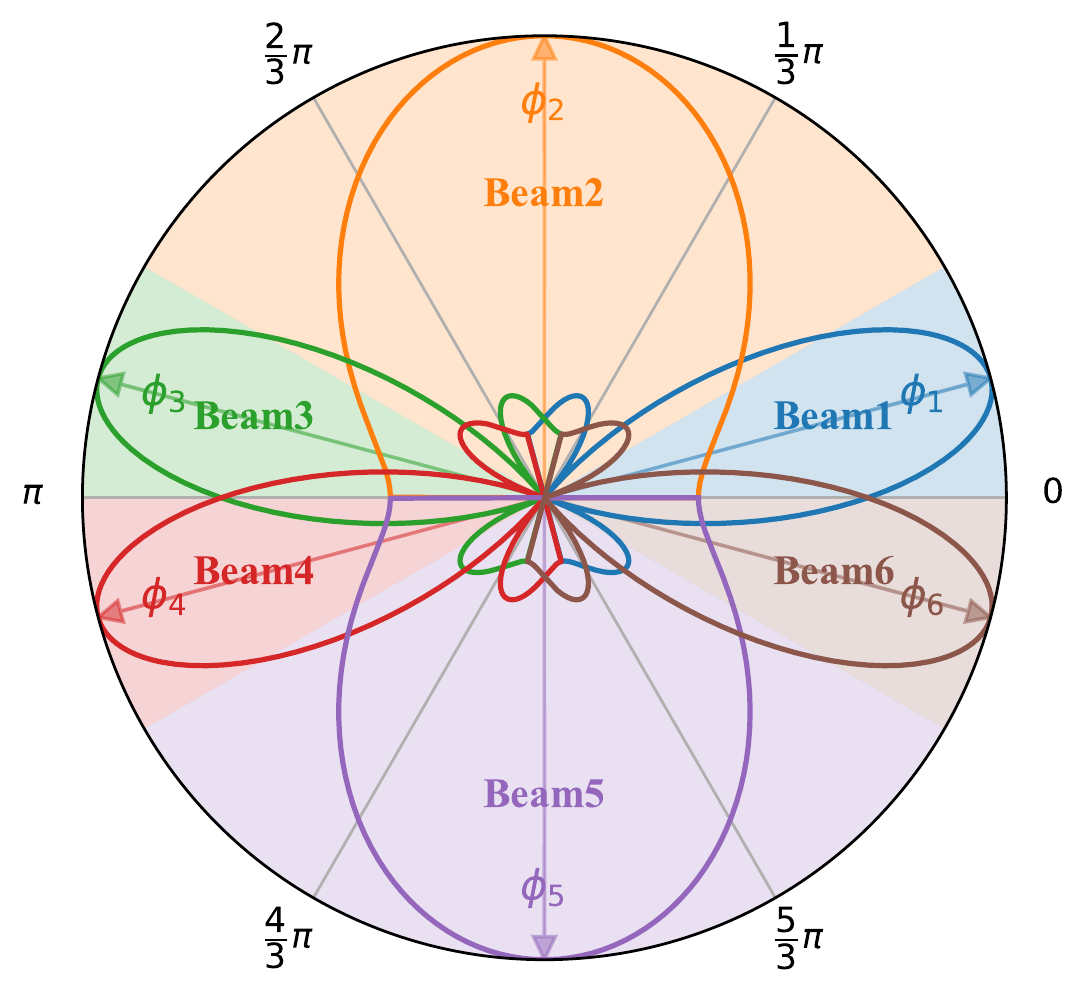}\label{subfig:model.action1}} & temp \\
    \subfloat[Action $a_{2}$.]{\includegraphics[width=0.8\mymultifigwidth]{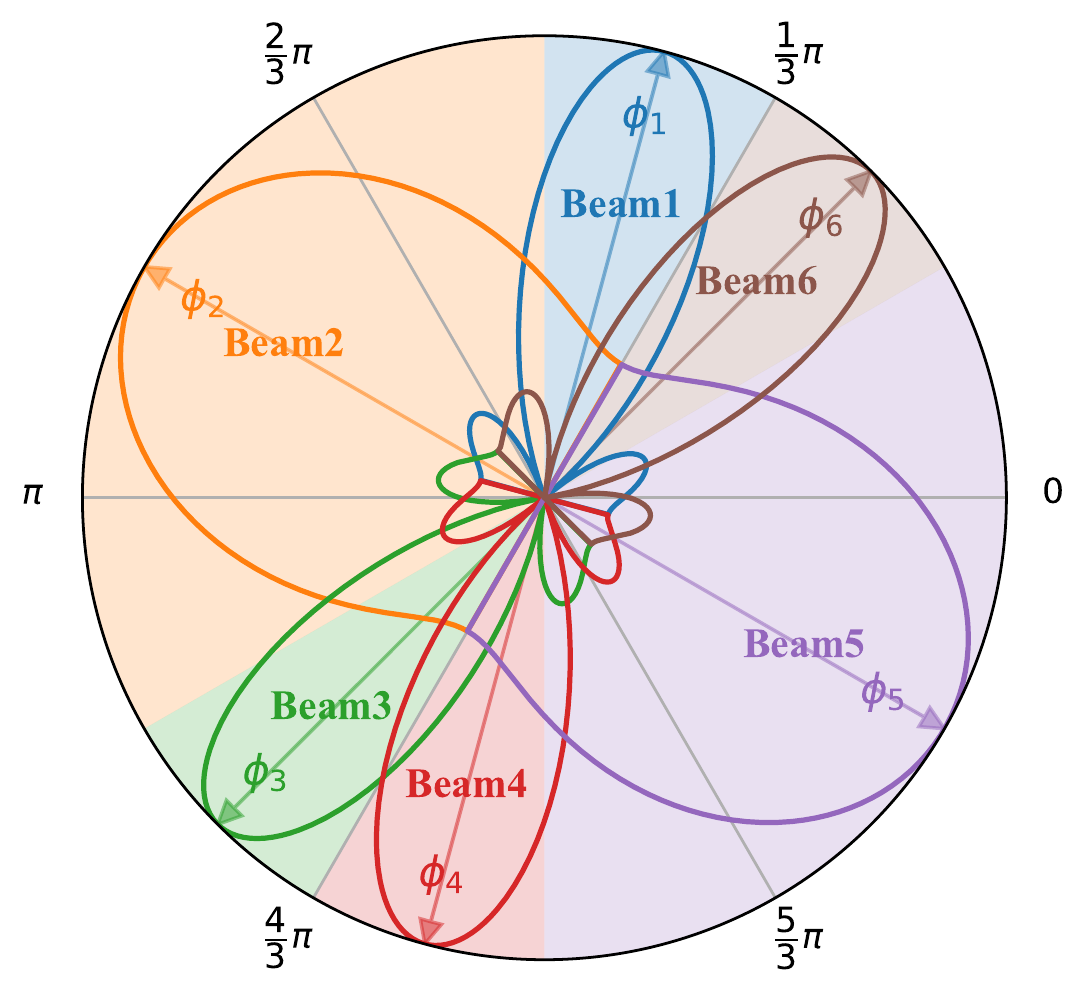}\label{subfig:model.action2}} & \subfloat[Action $a_{3}$.]{\includegraphics[width=0.8\mymultifigwidth]{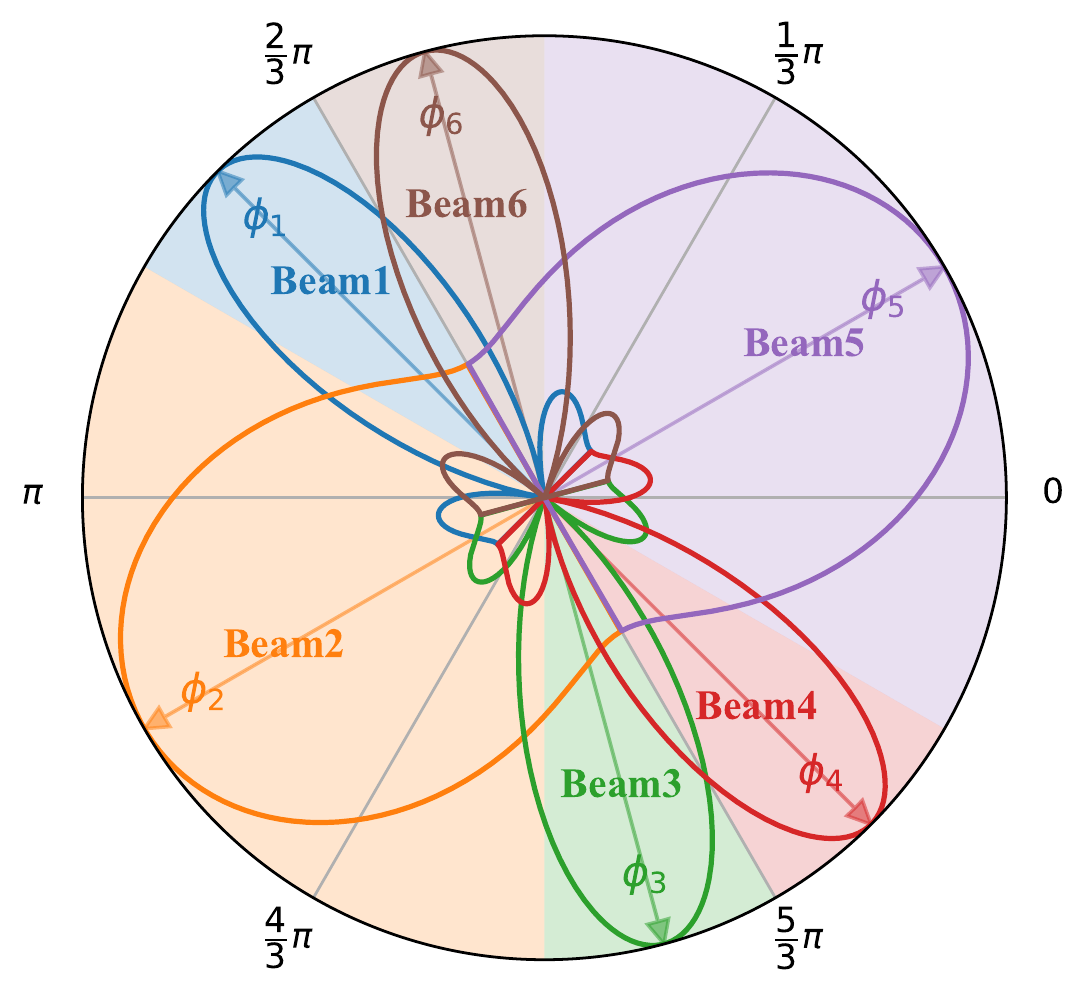}\label{subfig:model.action3}}
\end{tblr}
\caption{Example of dynamic beam-based random access.}
\label{fig:model.beams}
\end{figure*}

The action $a \in \mathcal{A}$ defines how the base station to adjust the direction and width of the beams, which can be expressed by
\begin{align}
\mathcal{A}=\left\{a \mid a = (\phi_{1}, \Theta_{1}, \phi_{2}, \Theta_{2}, \dotsc, \phi_{N_{\mathrm{g}}}, \Theta_{N_{\mathrm{g}}}) \right\},
\label{eq:model._action}
\end{align}
where $\phi_{i}$ and $\Theta_{i}$, $i \in \{1, 2, \dotsc, N_{\mathrm{g}}\}$ are the maximum gain direction and beamwidth of the $i$-th beam, respectively. It is noted that the direction and beamwidth of each beam strictly follow the requirements in~\eqref{eq:beam_constraint}. For the simplicity of illustration,  the action only includes several typical selections with the given number of beams, i.e.,
\begin{align}
\mathcal{A}=\left\{a_{1}, a_{2}, \dotsc, a_{\lvert \mathcal{A} \rvert} \right\},
\label{eq:model.action}
\end{align}
where $\lvert \mathcal{A} \rvert$ represents the size of the action space, and $a_{i},i \in \{1, 2, \dotsc, \lvert \mathcal{A} \rvert\}$ are pre-defined typical actions\footnote{Here a typical group of beams with unequal beamwidth is presented only as an illustration of beam-based access schemes, since the beam design is beyond the scope of this paper.}. Assume that the number of beams $N_{\mathrm{g}}$ is equal to $6$, and the action space $\mathcal{A}$ contains $3$ typical actions as shown in Table~\ref{tab:model.action}, then the beam gain of each action is given in Fig.~\ref{fig:model.beams}.

\begin{remark}
When one action is selected, e.g., $a_{1}$, the signal gains received by the base station from different directions may vary considerably, since multiple beams have different beamwidths. In other words, the base station have better signal gain but less coverage using one of beams, and vice versa. The beams used in actions $a_{2}$ and $a_{2}$ are obtained by rotating the beams of action $a_{1}$ counterclockwise by $\pi/3$ and $2\pi/3$, respectively. Therefore, the beams corresponding to different actions may better serve \ac{M2M} devices located in various directions.
\end{remark}


\subsection{State Transition}
\label{subsubsec:model.state_transition}

In this paper, the state transition is simply expressed as $s \xrightarrow{a} s'$, which means that the current state $s$ transfers to the new state $s'$ after executing action $a$. To better reflect the dynamic feature of random access process, the \ac{MDP} model of this paper adopts \textbf{\textit{episodic task}} based state transition process, including a series of relatively independent \textbf{\textit{episodes}}~\cite{sutton_reinforcement_1998}. \textit{Each episode} contains the following state transition steps, i.e.,

\subsubsection{\textbf{Phase 1}}
\label{itm:model.state_transition.1}

At the beginning of each episode, the \ac{M2M} devices in each sector initiate the random access process at the slot $t=0$ with the given arrival rates $\lambda_{j}$, ${j \in \{1, 2, \dotsc, N_{\mathrm{s} }\}}$. Correspondingly, the initial state is represented by $s = S_{(0)}$.

\subsubsection{\textbf{Phase 2}}
\label{itm:model.state_transition.2}

Based on the number of devices, the base station adjusts the direction and beamwidth according to the policy $\pi$, namely $a = \pi(s)$. Then, the base station operates the proposed beam-based random access protocol as described in Section \ref{sec:model}. Additionally, the \ac{M2M} devices that fail to access will continue to initiate an access request at the next slot $t'$. At this point, the state of the model is updated as $s' = S_{(t')}$.

\subsubsection{\textbf{Phase 3}}
\label{itm:model.state_transition.3}

If most of the \ac{M2M} devices initiating the random access process in the initial state $S_{(0)}$ have successfully connected the network before updating the new state $S_{(t')}$, the current episode terminates. This also means that only a few \ac{M2M} devices fail to access the network, i.e., $\sum_{j=1}^{N_{\mathrm{s}}} n_{j}' \leq n_{\mathrm{T}}$, where $n_{ \mathrm{T}}$ is the performance threshold. Otherwise, update the slot $t \leftarrow t'$ and state $s \leftarrow s'$, and repeat the \textbf{\textit{Phase~2}} of the state transition.

Finally, when an episode is terminated, the model transfers to a new independent episode and performs \textbf{\textit{Phase~1--3}} as described above.


\subsection{Reward}
\label{subsubsec:model.reward}


The reward $r(s,a)$ is a real-valued function of the state $s$ and the action $a$, reflecting the benefits and costs of the current instantaneous decision $a = \pi(s)$. Therefore, the reward definition has to be consistent with the optimization objective. Since the model of this paper relies on instantaneous value rather than statistical one, the successful probability of access can not be directly used as the optimization objective. To this end, an alternative optimization is required. Generally, the policy $\pi$ can be regarded as working well, if the access requests of all \ac{M2M} devices can be performed successfully and quickly. Meanwhile, it also implies that this policy $\pi$ can improve the successful probability of access. Therefore, we choose the access delay as the performance metric of random access rather than the probability. In conclusion, the goal of the model is to minimize the average delay of random access.

According to the above state transition process, when the model in a certain episode transfers from the current state $s = (n_{1}, n_{2}, \dotsc, n_{ N_{\mathrm{s}}})$ to a new state $s' = (n_{1}', n_{2}', \dotsc, n_{N_{\mathrm{s}}}')$ by taking the action $a$, the number of \ac{M2M} devices that fail to access the network can be expressed by
\begin{align}
\hat{r}(s,a) = \sum_{j=1}^{N_{\mathrm{s}}} n_{j}'.
\end{align}
Since these \ac{M2M} devices may try to access again in the new state $s'$, each of them spends one more time slot compared with those \ac{M2M} devices that have successfully accessed in the state $s$. Thus, $\hat{r}(s,a)$ can be the proxy of the ``\textit{access delay}'' during the state transition $s \xrightarrow{a} s'$, it also is the number of time slots used for random access. By summing up $\hat{r}(s,a)$ of all states in an episode, the ``\textit{total access delay}'' of \ac{M2M} devices can be obtained by
\begin{align}
\hat{G}_{\tau}
	= \sum_{t=0}^{T_{\tau}} \hat{r}\left(S_{(t)}, A_{(t)}\right) 
	= \sum_{m=1}^{\vert S_{(0)} \vert} \sum_{t=0}^{T_{\tau}} \mathbf{1}_{m \notin S_{(t)}},
\label{eq:model.sum_delay}
\end{align}
where $T_{\tau}$ represents the termination moment of the episode $\tau$, while $S_{(t)}$ and $A_{(t)}$ represent the state and action at the time slot $t$ of random access, respectively. $\vert S_{(0)} \vert$ is the number of \ac{M2M} devices that initiate random access requests in the initial state, namely $\vert S_{(0)} \vert = \sum_{j=1}^{ N_{\mathrm{s}}} n_{j}$. Let $\mathbf{1}_{m \notin S_{(t)}}$ indicate whether the $m$-th \ac{M2M} device in the initial state has successfully accessed the network at the time slot $t$. If the \ac{M2M} device has not yet connected to the network, i.e., $\mathbf{1}_{m \notin S_{(t)}} = 1$; otherwise, $\mathbf{1}_{m \notin S_{(t) }} = 0$. Thus, the access delay of the $m$-th \ac{M2M} device can be calculated by $\sum_{t=0}^{T_{\tau}} \mathbf{1}_{i \notin S_{(t)}}$. Next, the average access delay can be given by
\begin{align}
\bar{G}_{\tau} =\hat{G}_{\tau} / (\vert S_{(0)} \vert - \hat{r}(S_{(T_{\tau})}, A_{(T_{\tau})})),
\label{eq:model.av_delay}
\end{align}
where $\hat{r}(S_{(T_{\tau})} , A_{(T_{\tau})}) < n_{\mathrm{T}}$.  

Based on the above definition, the reward finally can be expressed by
\begin{align}
r(s,a) = -\hat{r}(s,a).
\label{eq:model.reward}
\end{align}
Here, a negative sign is adopted, this is because the reward definition of \ac{MDP} must satisfy the following assumption, i.e., the larger the reward value, the better the action. Accordingly, the cumulative reward from time slot $t$ to the termination moment of episode $\tau$ can be given by 
\begin{align}
G_{\tau(t)} &= \sum_{k=t}^{T_{\tau}} \gamma^{k-t} r(S_{(k)}, A_{(k)}) \notag \\
&= - \sum_{k=t}^{T_{\tau}} \gamma^{k-t} \hat{r}(S_{(k)}, A_{(k)}),
\label{eq:model.return}
\end{align}
where $\gamma$ is the so-called discount factor. Since the formulated model in this section is based on the episodic task, then $\gamma = 1$. In conclusion, maximizing the the expected cumulative reward $\mathbb{E}_{\pi}[G_{\tau(t)}]$ is equivalent to minimizing the average ``access delay'' of \ac{M2M} devices, which is consistent with our original optimization goal.

\section{Dynamic Random Access Scheme with \acs*{DDQN}}
\label{sec:solution} 

\begin{algorithm}[!t]
\ifCLASSOPTIONonecolumn
    \small
\fi
\setlength{\algowidth}{\hsize}
\setlength{\hsize}{\linewidth}
\caption{Dynamic random access scheme for non-uniform grouping case based on \acs*{DDQN}.}
\label{alg:model.DDQN}
\setlength{\hsize}{\algowidth}
\setlength{\algowidth}{\linewidth}
\KwIn{}
Initialize the prediction network $Q(s,a;\myvec{\theta})$ with random weights $\myvec{\theta}$\;
Initialize the target network $Q(s,a;\myvec{\theta}^-)$ with weights $\myvec{\theta}^- = \myvec{\theta}$\;
Initialize the replay memory $D$\;
Initialize $\varepsilon = 1$.
\BlankLine
\Kwflow{}
\For{$\tau \leftarrow 1,\dots,\infty$}{
	{Initialize the state $s=S_{(0)}$ according to the arrival rate $\lambda_{j}$, ${j \in \{1, 2, \dotsc, N_{\mathrm{s}}\}}$}\;
	\For{$t \leftarrow 1,\dots, \infty$}{
		{Generate a random number $p$, $0 < p \leq 1$} \tcc*[r]{$\varepsilon-$greedy}
		{$a \leftarrow
			\begin{cases}
			a \in_{\mathrm{R}} \mathcal{A} & 0 \leq p <  \varepsilon \\
			\argmax_{a \in \mathcal{A}} Q(s,a;\myvec{\theta}) & \varepsilon \leq p < 1 \\
			\end{cases}$
		}\;
		{Take the action $a$ and perform state transition $s \xrightarrow{a,r(s,a)} s'$}\;
		{Save the experience $e\leftarrow(s,a,r(s,a),s')$ into the replay memory $D$}\;
		{Random sample a batch of experiences $\tilde{D}$ from $D$}\;
		{Calculate $\operatorname{L}(\myvec{\theta})$} \tcc*[r]{according to~\eqref{eq:model.loss_function}}
		{Update weights $\myvec{\theta}$ by optimizer}\;
		\If(\tcc*[f]{exponential decay of $\varepsilon$}) {$\varepsilon > \varepsilon_{\mathrm{min}}$} {
			{$\varepsilon \leftarrow \varepsilon \cdot \varepsilon_{\mathrm{decay}}$}\;
		}
		\If(\tcc*[f]{every $n_{\myvec{\theta}}$ steps}) {$t \mod n_{\myvec{\theta}} = 0$} {
			{$\myvec{\theta}^- \leftarrow \myvec{\theta}$}\;
		}
		\If(\tcc*[f]{episode end $T_{\tau} = t$}) {$\hat{r}(s, a) \leq n_{T}$} {
			\textbf{break}\;
		}
		{$s \leftarrow s'$\;}
	}
}
\BlankLine
\KwOut{Obtain the optimal policy $\pi^*$ based on~\eqref{optpol}.}
\end{algorithm}

\begin{table*}[!t]
\centering
\caption{Main Parameters in Simulations}
\label{tab:model.experiment_configuration}
\begin{tblr}{
    width = 0.8\linewidth,
    colspec = {X[1,c,m]X[4,l,m]X[2,l,m]},
    hlines,
    hline{2} = {1}{-}{},
    hline{2} = {2}{-}{},
    hline{3} = {1}{-}{},
    hline{3} = {2}{-}{},
    hline{14} = {1}{-}{},
    hline{14} = {2}{-}{},
    hline{15} = {1}{-}{},
    hline{15} = {2}{-}{},
    vline{2-3},
    row{1} = {font=\bfseries},
    row{2} = {font=\bfseries},
    row{14} = {font=\bfseries},
    cell{2}{1} = {c=3}{c,m},
    cell{14}{1} = {c=3}{c,m},
}
    Parameter & Description & Value \\
    \ac{MDP} Model Related Parameters & temp & temp \\
    $\lambda$ & Total arrival rate of M2M devices  & $150$, $300$ \\
	$\rho$ & Distribution ratio & $2$, $5$, $20$ \\
	$d$ & Distance between M2M devices and base station  & $0$ - $10$ \myunit{km} \\
	$N_{\mathrm{p}}$ & Number of the preambles & $48$ \\
	$N_{\mathrm{g}}$ & Number of beams & $6$ \\
	$P_{\mathrm{t}}$ & Transmit power of M2M device & $23$ \myunit{dBm} \\
	$G_{\mathrm{t}}$ & Transmit antenna gain of M2M device & $0$ \myunit{dBi} \\
	$G_{\mathrm{r}}$ & Receive antenna gain of base station  & $18$ \myunit{dBi} \\
	$\sigma$ & Standard variance of shadowing  & $8$ \myunit{dB}  \\
	$\operatorname{PL}(d)$ & Path loss & $120.9 + 37.6 \log_{10}(d)$ \\
	$\Gamma$ & Demodulation threshold & $-110$ \myunit{dBm} \\
	\ac{DDQN} Algorithm Related Parameters & temp & temp \\
	$\varepsilon_{\mathrm{min}}$ & Minimum $\varepsilon$  & $0.01$ \\
	$\varepsilon_{\mathrm{decay}}$ & $\varepsilon$ Exponential decay & $0.99$ \\
	$\alpha$ & Learning rate & $0.001$ \\
	$\lvert D \rvert$ &  Size of replay memory $D$& $\num{1200}$ \\
	$\lvert \tilde{D} \rvert$ & Number of samples per batch & $64$ \\
	$n_{\myvec{\theta}}$ &  Weight update period of the target networks $\myvec{\theta}^-$ & $16$ \\
\end{tblr}
\end{table*}

To find the optimal policy $\pi^*$ for maximizing the expected cumulative reward  $\mathbb{E}_{\pi^*}[G_{\tau(t)}]$, we propose an dynamic random access scheme based on the \ac{DDQN} algorithm to solve the \ac{MDP} model. The \ac{DDQN} algorithm is a value-based reinforcement learning technique that utilizes a neural network to approximate the value function $Q_{\pi} (s,a)$, i.e.,
\begin{align}
&\phantom{=}Q(s,a; \myvec{\theta}) \approx Q_{\pi}(s,a) \notag \\
&=\mathbb{E}_{\pi}\left[G_{\tau(t)} \mid S_{(t)}=s, A_{(t)}=a \right] \notag \\
&=\mathbb{E}_{\pi}\left[\sum_{k=t}^{\infty} \gamma^{k-t} r(S_{(k)}, A_{(k)}) \big\vert S_{(t)}=s, A_{(t)}=a \right],
\end{align}
where $\myvec{\theta}$ is the weight parameter of the neural network $Q(s,a; \myvec{\theta})$. $Q(s,a; \myvec{\theta})$ is also known as the \textbf{\textit{prediction network}}, whose input is a $N_{\mathrm{s}}$-dimension vector representing the current state $s$ and output is a $\lvert \mathcal{A} \rvert$-dimension vector representing the action value corresponding to each action $a$. Then, by using the action value function, the corresponding policy can be directly obtained by $\pi(s) = \argmax_{a \in \mathcal{A}} Q_{\pi}(s, a)\approx \argmax_{a \in \mathcal{A}} Q(s, a; \myvec{\theta})$.

In order to estimate the action value function accurately, we have to use another \textbf{\textit{target network}}, namely $Q(s,a; \myvec{\theta}^-)$, which has the same structure with the prediction network~\cite{hasselt_deep_2016}. Then, the \ac{MSE} function can be defined by
\begin{align}
\operatorname{L}(\myvec{\theta}) &= \frac{1}{\lvert \tilde{D} \rvert} \sum_{e \in \tilde{D}} \Big[\big(r(s,a) - Q(s,a; \myvec{\theta}) \notag  \\
&\qquad + \gamma Q(s', \argmax_{a'}{Q(s',a'; \myvec{\theta})};\myvec{\theta}^-) \big)^2 \Big],
\label{eq:model.loss_function}
\end{align}
where $e=(s, a, r(s, a), s')$ is an experience sample including the state $s$, action $a$, new state $s'$ and the resulting reward $r(s, a)$ in the state transition process of $s \xrightarrow{a, r(s,a)} s'$. $\tilde{D}$ is a set of $\vert \tilde{D} \vert$ experiences, which are selected from the replay memory $D$ with a \ac{FIFO} queue. The \ac{DDQN} algorithm updates the weight parameters of the neural network iteratively with the objective of minimizing the loss function $\operatorname{L}(\myvec{\theta})$ until the optimal action value function is estimated, i.e., $Q(s, a; \myvec{\theta}^*) \approx Q_{\pi^*}(s, a)$. Afterwards, the optimal policy can be written as
\begin{align}
\label{optpol}
\pi^*(s) &= \argmax_{a \in \mathcal{A}} Q_{\pi^*}(s, a) \notag \\
&\approx \argmax_{a \in \mathcal{A}} Q(s, a; \myvec{\theta}^*).
\end{align}

Algorithm~\ref{alg:model.DDQN} proposes a dynamic random access scheme for non-uniform grouping case based on \acs*{DDQN}. First of all, the related model parameters including the so-called prediction network, target network, and replay memory are initialized. Then, the iteration of a episode is carried out. At the beginning of each episode, the initial state $s=S_{(0)}$ is generated according to the arrival rates of access in each sector, i.e., $\lambda_{j}$, $j \in \{1, 2, \dotsc, N_{\mathrm{s}}\} $. Subsequently, during each episode, the action $a$ is selected based on the $\varepsilon$-greedy rule in the iteration of each access slot. In particular, the greedy policy is selected with the probability of $1 - \varepsilon$, while a random action is chosen with the probability $\varepsilon$. The $\varepsilon$ decays exponentially with the parameter of $\varepsilon_{\mathrm{decay}}$ at each iteration until it decays to the minimum value of $\varepsilon_{\mathrm{min}}$. The $\varepsilon$-greedy rule can efficiently prevent the algorithm from falling into the local optimal solution. Next, the algorithm executes the action $a$ and performs the state transition. Meanwhile, the generated experience $e$ are saved in the replay memory $D$. Finally, a batch of $\tilde{D}$ experiences are randomly selected from the replay memory $D$ to calculate the loss function $\operatorname{L}(\myvec{\theta})$, and the weight parameter $\myvec{\theta}$ of the prediction network is updated. It is noticed that only after $n_{\myvec{\theta}}$ iterations, the weight parameter $\myvec{\theta}^-$ of the target network is updated with the weight parameter $\myvec{\theta}$ of the prediction network, namely $\myvec{\theta}^- \leftarrow \myvec{\theta}$. When the algorithm converges, the optimal action value function $Q(s, a; \myvec{\theta}^*) \approx Q_{\pi^*}(s, a)$ is obtained as well as the optimal policy $ \pi^*(s) \approx \argmax_{a \in \mathcal{A}} Q(s, a; \myvec{\theta}^*)$.

\section{Simulation Results and Analysis}
\label{sec:simulation}

\subsection{Simulation Configuration}

In the simulations, a base station with massive \ac{MIMO} is divided into $N_{\mathrm{s}}=6$ sectors. The sectors are bounded by the directions of $[(j-1)\pi/3 - \pi/6, (j-1)\pi/3 + \pi/6], j \in \{1, 2, \dotsc, N_{\mathrm{s}}\}$. The distance between the \ac{M2M} devices and the base station varies randomly from $0$ to $10$ \myunit{km}. Moreover, the base station generates $N_{\mathrm{g}}=6$ beams, and adjusts the direction and width of the beams according to the action space $\mathcal{A}$ as shown in Table~\ref{tab:model.action}. To reflect the non-uniform feature of device distribution, one of sectors is set as the high-density area with the high arrival rate of access, such as $\lambda_{1} = \lambda_\mathrm{high}$, while other sectors are set with $\lambda_{j} = \lambda_{\mathrm{low}}, j=2, 3, \dots, N_s$. The total arrival rate is defined as $\lambda = \sum_{j=1}^{N_{\mathrm{s}}} \lambda_j$. Finally, we also define a distribution ratio between the high-density and low-density sectors as $\rho =  \frac{ \lambda_{\mathrm{high}} } {(N_s - 1) \lambda_{\mathrm{low}}}$ to evaluate the performance under various arrival rate. The main parameters of simulations are shown in Table~\ref{tab:model.experiment_configuration}.

In the \ac{DDQN} algorithm, both the prediction and the target networks use a two-layer fully connected neural network, in which each fully connected layer contains $64$ neurons with the \ac{ReLU} activation function. The size of the replay memory $\lvert D \rvert$ is set to $\num{1200}$. During each iteration, the algorithm selects $64$ experiences per batch from the replay memory to calculate the loss function $\operatorname{L}(\myvec{\theta})$ . In addition, the Adam optimizer with a learning rate of $\alpha = 0.001$ is applied to update the weight parameters $\myvec{\theta}$ of the prediction network. Finally, the parameters of the $\varepsilon$-greedy rule are set as $\varepsilon_{\mathrm{min}}=0.01$ and $\varepsilon_{\mathrm{decay}}=0.99$, respectively. The weight update period of the target network $\myvec{\theta}^-$ is set as $ n_{\myvec{\theta}}=16$.

\subsection{Performance Analysis}

\subsubsection{Training Performance}

\begin{figure}[!t]
	\centering
	\subfloat[Loss function value $\operatorname{L}(\myvec{\theta})$ versus episode]{\includegraphics[width=\mymultifigwidth]{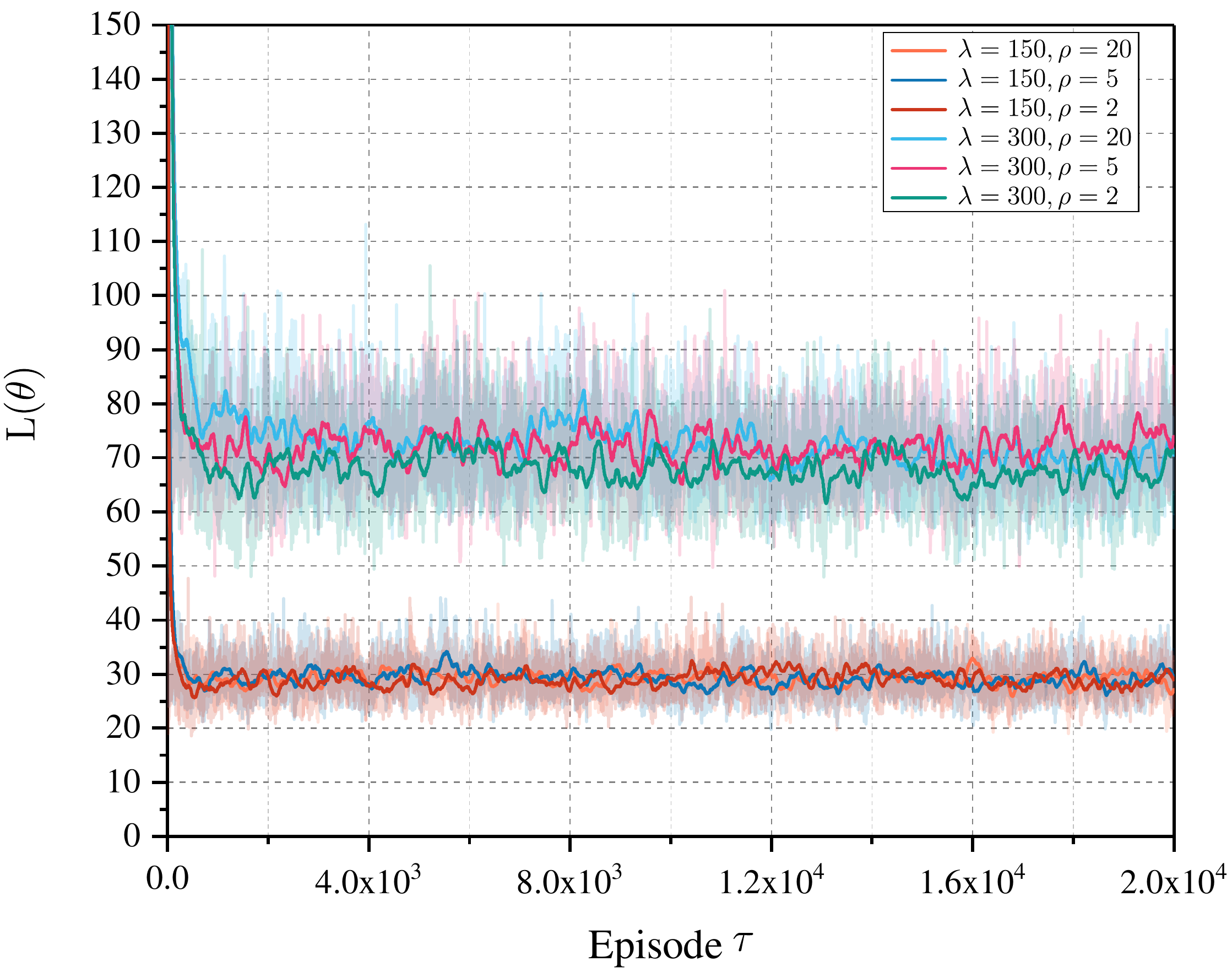}\label{subfig:loss}}
	\hfill
	\subfloat[Average action value versus episode]{\includegraphics[width=\mymultifigwidth]{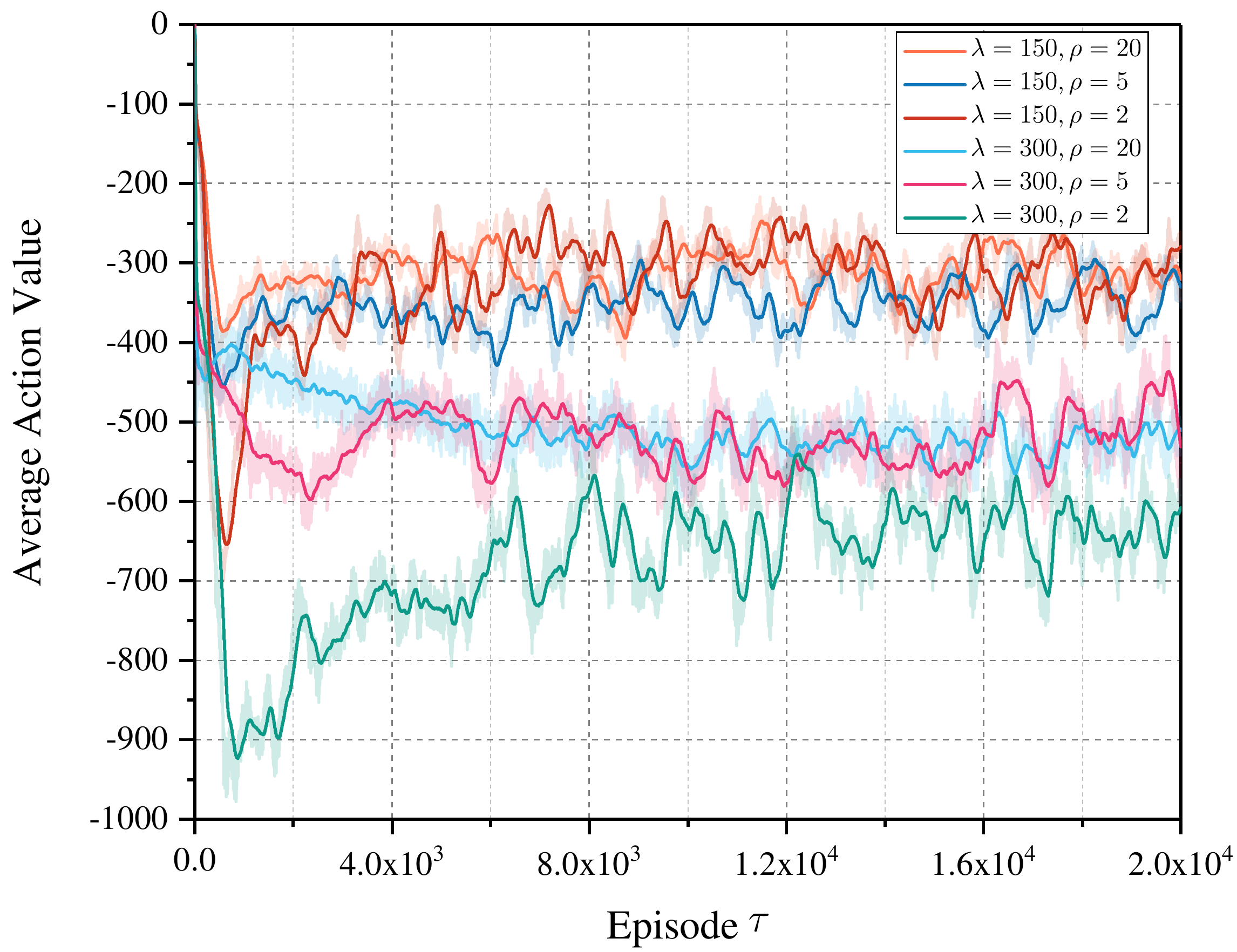}\label{subfig:avg_q}}
	\caption{Training performance of the proposed \acs*{DDQN}-based dynamic random access algorithm.}
	\label{fig:model.training}
\end{figure}

\renewcommand\TblrOverlap[1]{#1}
\begin{table*}[!t]
\centering
\caption{Average Access Delay of \acs*{M2M} Devices}
\label{tab:model.mean_delay}
\begin{talltblr}[
    label = none,
    note{1} = {The gain is defined as the average access delay of the \textbf{DDQN-BU} scheme divided by that of the \textbf{Static-BE} scheme.},
]{
    width = 0.8\linewidth,
    colspec = {X[1,l,m]X[1,c,m]X[1,c,m]X[1,c,m]X[1,c,m]},
    hlines,
    hline{3} = {1}{-}{},
    hline{3} = {2}{-}{},
    vline{2-5},
    row{1} = {font=\bfseries},
    cell{1}{1} = {r=2}{c,m},
    cell{1}{4} = {c=2}{c,m},
}
    Case & Static-BE & Random-BU & DDQN-BU & temp \\
    temp & {Average Value \\ (Time slots)} & {Average Value \\ (Time slots)} & {Average Value \\ (Time slots)} & Gain\TblrNote{1} \\
    $\lambda=150, \rho=2$ & $3.72$ & $3.56$ & $3.42$ & $4.1 \%$ \\
	$\lambda=150, \rho=5$ & $4.01$ & $3.94$ & $3.53$ & $10.4 \%$ \\
	$\lambda=150, \rho=20$ & $4.29$ & $4.28$ & $3.66$ & $14.7 \%$ \\
	$\lambda=300, \rho=2$ & $4.66$ & $4.65$ & $4.15$ & $10.7 \%$ \\
	$\lambda=300, \rho=5$ & $5.24$ & $5.32$ & $4.44$ & $16.6 \%$ \\
	$\lambda=300, \rho=20$ & $5.96$ & $6.24$ & $4.75$ & $23.9 \%$
\end{talltblr}
\end{table*}

Fig.~\ref{fig:model.training} shows the training performance of the proposed \ac{DDQN}-based scheme after iterating $2 \times 10^4$ episodes in terms of the loss function and average action value. All solid lines represent the results after \ac{EMA} processing with a weight of $0.99$, while the original data is given in the shadow parts. The values of loss function under different arrival rates and distribution ratios are first presented in  Fig.~\subref*{subfig:loss}. Each curve shows a rapid downward trend during the first $10^3$ episodes, and then is stabilized within a certain range. Particularly, the loss function value in the case of $\lambda = 300$ ends up fluctuating between $60$ - $80$, while the loss function value under $\lambda = 150$ remains stable at about $30$. After about the $2 \times 10^3$ episode, each loss function almost converges.

Subsequently, Fig.~\subref*{subfig:avg_q} gives the performance of average action value. Here the average action value refers to the average output of the prediction network in each episode, i.e., $\frac{1}{T_{\tau} \cdot \vert \mathcal{A} \vert} \sum_{t=1 }^{T_{\tau}} \sum_{a \in \mathcal{A}} Q(S_{(t)}, S_{(t)}; \myvec{\theta})$. Due to the different trajectories of state transition, the instantaneous action value is not comparable, and it cannot accurately reflect the convergence of the algorithm. Thus, instead of the instantaneous action value, the average one is chosen as the performance metric.

In the first $10^3$ episodes of Fig.~\subref*{subfig:avg_q}, each curve starts at $0$ and shows a rapid downward trend. This is because the weight parameters of the prediction network are initialized with the standard \textit{Glorot}, resulting in a small output value at the beginning of the training. Then, as the weight parameters are updated rapidly during the training progress, the average action value of the prediction network changes dynamically. After the $10^3$ episodes, the curves under various $\lambda $ show different trends. Specifically, in the case of $\lambda = 150$, three curves first rise slowly and then fluctuate between $-250$ and $-400$. On the other hand, in the case of $\lambda = 300$, the curves with $\rho= 20$ or $5$ still decline slowly, and gradually stabilize at around $-500$. Furthermore, when $\lambda = 300$ and $\rho = 2$, the corresponding curve eventually stabilizes between $-600$ and $-700$. Lastly, all of them almost reach a state of convergence after $10^4$ episodes. In addition, it is noted that the average action value in the case of $\lambda = 300$ is much lower than that of $\lambda = 150$. This is because the larger arrival rate $\lambda$, the greater chance of collision, which leads to a larger access delay and a lower action value. According to Fig.~\ref{fig:model.training}, it is evident that the proposed scheme with the \ac{DDQN} algorithm becomes convergent fast and steadily.

\subsubsection{Delay Performance}

The proposed scheme based on multiple \underline{\textbf{b}}eams with \underline{\textbf{u}}nequal beamwidth (hereinafter referred to as \textbf{DDQN-BU}) is compared with the following random access schemes, i.e.,
\begin{itemize}
\item Static scheme based on multiple \underline{\textbf{b}}eams with \underline{\textbf{e}}qual beamwidth (\textbf{Static-BE}): In this scheme, all beams have the equal beamwidth, i.e., $\Theta_{i} = 2\pi / N_{\mathrm{g}}$. Thus, there is no need to make selection during the random access~\cite{xiong_group-based_2018}.

\item Random scheme based on multiple \underline{\textbf{b}}eams with \underline{\textbf{u}}nequal beamwidth (\textbf{Random-BU}): This scheme randomly selects one of the beams in Fig.~\ref{fig:model.beams} to serve all \ac{M2M} devices, regardless of their spatial distribution.
\end{itemize}

Table~\ref{tab:model.mean_delay} gives the average ``\textit{access delay}'' when different random access schemes are applied under various arrival rates $\lambda$ and distribution ratios $\rho$. It can be seen that the \textbf{DDQN-BU} scheme has the lower average access delay than others in all cases. At the same time, its relative gain to \textbf{Static-BE} scheme increases with the higher arrival rate and the larger distribution ratio. For example, in the case of $\lambda = 300$ and $\rho=20$, the \textbf{DDQN-BU} scheme has the lowest average access delay of $4.75$ slots, while the average access delay of the \textbf{Static-BE} scheme and \textbf{Random-BU} scheme are $5.96$ and $6.24$ slots, respectively. Therefore, the \textbf{DDQN-BU} scheme allows \ac{M2M} devices to access the network faster by dynamically adjusting the beam direction and beamwidth, and owns optimal access performance.

When the distribution ratio $\rho$ is first fixed, the average access delay of each scheme becomes larger with the increase of the arrival rate $\lambda$. We can further find that the gain of the \textbf{DDQN-BU} scheme compared with the \textbf{Static-BE} scheme also become more intuitive with the increase of the arrival rate $\lambda$. For example, at $\rho=20$, the gain of \textbf{DDQN-BU} at $\lambda = 150$ and $\lambda = 300$ are $14.7 \%$ and $23.9 \%$, respectively. This shows that the larger arrival rate $\lambda$ is, the more intuitive optimization effect of the \textbf{DDQN-BU} scheme on the access performance achieves. 

Next, we fix the arrival rate $\lambda$ but increase the distribution ratio $\rho$. The average access delay of each scheme increases when the non-uniform distribution of devices becomes stronger, namely the larger $\rho$. This phenomenon is expected because \ac{M2M} devices collide more frequently as the number of devices increases within a given sector, and thus the average access delay eventually rises. More importantly, the gain of the \textbf{DDQN-BU} scheme also becomes larger with the increase of the distribution ratio $\rho$. For example, at $\lambda = 150$, the gain improvements are $4.1 \%$, $10.4 \%$ and $14.7\%$ for $\rho = 2$, $\rho = 5$ and $\rho = 20$, respectively.
This shows that the stronger non-uniform distribution, the more intuitive optimization effect of the \textbf{DDQN-BU} scheme on access performance. Compared with the reference schemes, the \textbf{DDQN-BU} scheme can better handle the access problem in the case of non-uniform distribution of \ac{M2M} devices.

\begin{figure}[!t]
	\centering
	\includegraphics[width=\mysinglefigwidth]{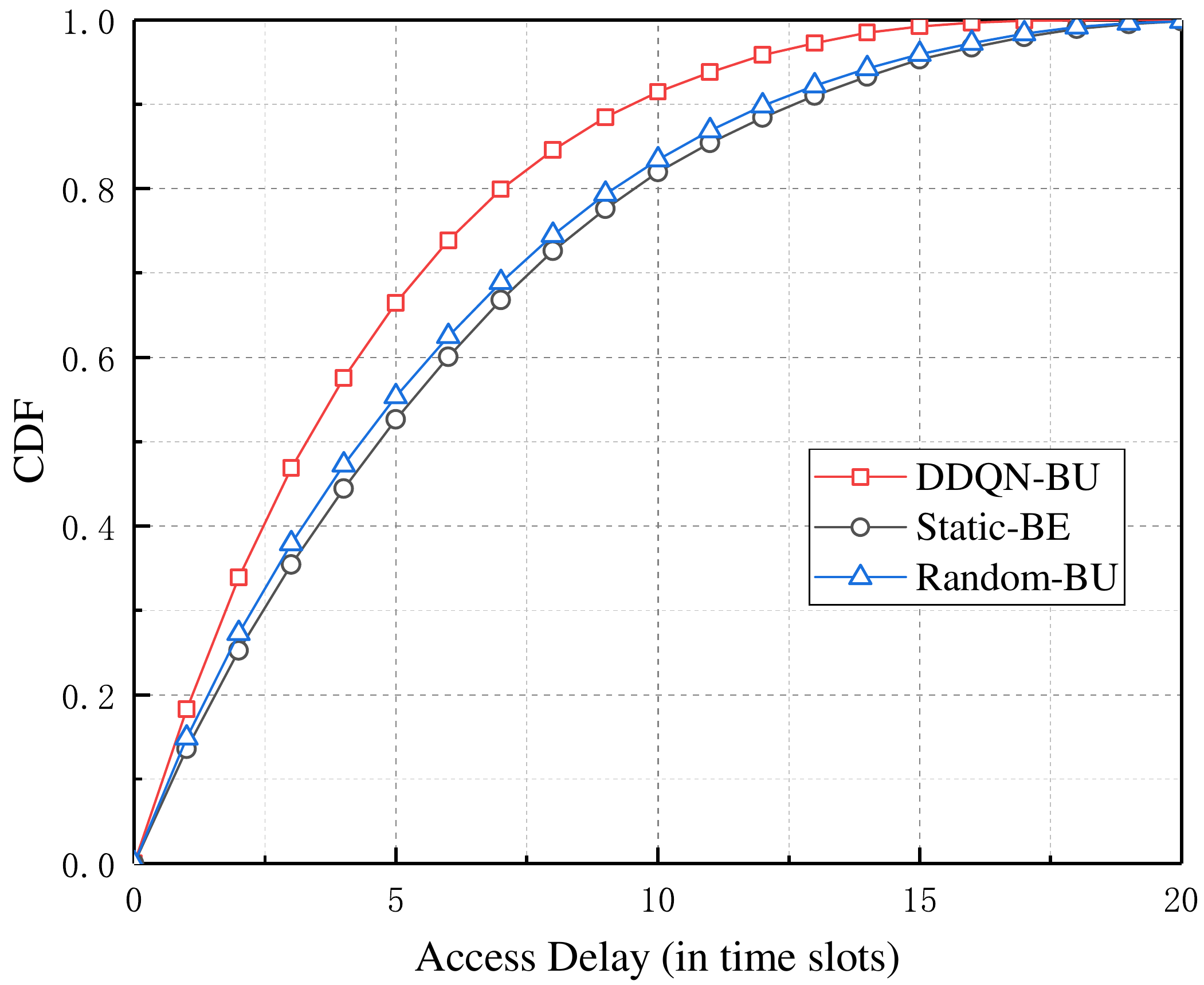}
	\caption{\acs*{CDF} performance of access delay under different access schemes with $\lambda=300 $ and $\rho=20$.}
	\label{fig:model.delay_cdf}
\end{figure}

Finally, we further analyze the \ac{CDF} performance of access delay. The \ac{CDF} curves of access delay have the same trend under different arrival rates $\lambda$ and distribution ratios $\rho$, thus here only the case of $\lambda=300$ and $\rho=20$ is taken as an example for the sake of analysis. As shown in Fig.~\ref{fig:model.delay_cdf}, when the access delay is small, all curves increase rapidly, such an increase becomes slower as the access delay becomes larger. It means that most \ac{M2M} devices have a low access delay, and only a few devices have a high access delay. For example, when the \textbf{DDQN-BU} scheme is adopted, about $66.4\%$, $91.5\%$ and $99.2\%$ of \ac{M2M} devices access the network successfully within $5$, $10$ and $15$ time slots, respectively. Furthermore, the \ac{CDF} curve of the \textbf{DDQN-BU} scheme rises the fastest among all ones. This shows that the \textbf{DDQN-BU} scheme dynamically adjusts the beam direction and beamwidth according to the distribution of devices, which can efficiently reduce the access delay and improve the overall access performance.

\section{Conclusion}
\label{sec:conclusion}

Many \ac{IoT} applications have the non-uniform distribution of devices in the space dimension, and the dynamic service arrival rate in the time dimension. As a result, this paper proposed a beam-based dynamic random access scheme in massive \ac{MIMO} systems. Specifically, the base station can dynamically adjust a group of beams with different bandwidths based on the distribution of devices, which can efficiently balance the number of \ac{M2M} devices served by different beams, reducing collisions during the access process. We also formulated this dynamic problem as a \ac{MDP} model, and proposed a dynamic beam-based random access scheme based on the \ac{DDQN} algorithm to obtain the optimal solution. Lastly, the performance of the proposed scheme were validated by simulations. The proposed scheme were proved to converge rapidly and stably. After the model is converged, the optimal policy can be determined by using the trained prediction neural network. The simulation results show that the proposed scheme performs much better than the other two reference schemes in terms of the average access delay.

\appendices

\section{Calculation of Beam Gain}
\label{sec:appendix}
 
For the sake of analysis, all beams are generated by the \ac{ULA}, and the beamwidth is set as the \ac{HPBW}~\cite{balanis_antenna_2016}, i.e.,
\begin{align}
\Theta_i = 2 \left[\frac{\pi}{2} - \cos^{-1} \left( \frac{1.391 \lambda_{\mathrm{a}}}{\pi N_{\mathrm{a}} d_{\mathrm{a}}} \right) \right], i \in \{1, 2, \dots, N_{\mathrm{g}} \},
\label{eq:HPBW}
\end{align}
where $\lambda_{\mathrm{a}}$ represents the wavelength, $d_{\mathrm{a}}$ represents the spacing of array elements, $N_{\mathrm{a}}$ represents the number of array elements, and $d_{\mathrm{a}} / \lambda_{\mathrm{a}} = 1/4$. Then, the number of array elements required to generate the beam with the beamwidth in~\eqref{eq:HPBW}  can be calculated by
\begin{align}
N_{\mathrm{a}} = \left\lceil \frac{1.391 \lambda_{\mathrm{a}}} {\pi d_{\mathrm{a}} \cos \left(\frac{\pi}{2} - \frac{\Theta_i}{2} \right)} \right\rceil.
\label{eq:antenna_number_of_HPBW}
\end{align}
Finally, the gain of the beam is given by 
\begin{align}
f_i(\theta) = 
\frac{\sin \left\{ \frac{N_{\mathrm{a}}}{2} \left[k d_{\mathrm{a}} \cos(\theta - \phi_i) + \beta \right] \right\}}
{N_{\mathrm{a}} \sin \left\{ \frac{1}{2} \left[k d_{\mathrm{a}} \cos(\theta - \phi_i) + \beta \right] \right\}},
\label{eq:beam_gain}
\end{align}
where  $k$ is equal to $2\pi / \lambda_{\mathrm{a}}$, and $\beta$ represents the phase difference between array elements. The maximum gain of the beam is set to be 1 for the sake of normalization, namely $\max f_i(\theta) = 1$.

%
\bibliographystyle{IEEEtran}
\bibliography{IEEEabrv,Bib/Ref}

\begin{thebibliography}{10}
\providecommand{\url}[1]{#1}
\csname url@samestyle\endcsname
\providecommand{\newblock}{\relax}
\providecommand{\bibinfo}[2]{#2}
\providecommand{\BIBentrySTDinterwordspacing}{\spaceskip=0pt\relax}
\providecommand{\BIBentryALTinterwordstretchfactor}{4}
\providecommand{\BIBentryALTinterwordspacing}{\spaceskip=\fontdimen2\font plus
\BIBentryALTinterwordstretchfactor\fontdimen3\font minus
  \fontdimen4\font\relax}
\providecommand{\BIBforeignlanguage}[2]{{%
\expandafter\ifx\csname l@#1\endcsname\relax
\typeout{** WARNING: IEEEtran.bst: No hyphenation pattern has been}%
\typeout{** loaded for the language `#1'. Using the pattern for}%
\typeout{** the default language instead.}%
\else
\language=\csname l@#1\endcsname
\fi
#2}}
\providecommand{\BIBdecl}{\relax}
\BIBdecl

\bibitem{Varsier2021}
N.~Varsier, L.-A. Dufrene, M.~Dumay, Q.~Lampin, and J.~Schwoerer, ``A {5G} new
  radio for balanced and mixed {IoT} use cases: {Challenges} and key enablers
  in {FR1} band,'' \emph{{IEEE} Commun. Mag.}, vol.~59, no.~4, pp. 82--87, Apr.
  2021.

\bibitem{Tataria2022}
H.~Tataria, M.~Shafi, M.~Dohler, and S.~Sun, ``Six critical challenges for {6G}
  wireless systems: {A} summary and some solutions,'' \emph{{IEEE} Veh.
  Technol. Mag.}, vol.~17, no.~1, pp. 16--26, Mar. 2022.

\bibitem{Tanab2021}
M.~El-Tanab and W.~Hamouda, ``An overview of uplink access techniques in
  machine-type communications,'' \emph{{IEEE} Netw.}, vol.~35, no.~3, pp.
  246--251, May/Jun. 2022.

\bibitem{laya_is_2014}
A.~Laya, L.~Alonso, and J.~Alonso-Zarate, ``\BIBforeignlanguage{en}{Is the
  random access channel of {LTE} and {LTE-A} suitable for {M2M} communications?
  {A} survey of alternatives},'' \emph{\BIBforeignlanguage{en}{{IEEE} Commun.
  Surveys Tuts.}}, vol.~16, no.~1, pp. 4--16, Firstquarter 2014.

\bibitem{ghavimi_m2m_2015}
F.~Ghavimi and H.-H. Chen, ``{M2M} communications in {3GPP} {LTE}/{LTE-A}
  networks: Architectures, service requirements, challenges, and
  applications,'' \emph{{IEEE} Commun. Surveys Tuts.}, vol.~17, no.~2, pp.
  525--549, Secondquarter 2015.

\bibitem{3GPP_tr37.868_2011}
``Study on {RAN} improvements for machine-type communications,'' 3GPP TR 37.868
  v11.0.0, Sep. 2011.

\bibitem{lien_cooperative_2012}
S.-Y. Lien, T.-H. Liau, C.-Y. Kao, and K.-C. Chen,
  ``\BIBforeignlanguage{en}{Cooperative access class barring for
  machine-to-machine communications},'' \emph{\BIBforeignlanguage{en}{{IEEE}
  Trans. Wireless Commun.}}, vol.~11, no.~1, pp. 27--32, Jan. 2012.

\bibitem{wang_optimal_2015}
Z.~Wang and V.~W.~S. Wong, ``\BIBforeignlanguage{en}{Optimal access class
  barring for stationary machine type communication devices with timing advance
  information},'' \emph{\BIBforeignlanguage{en}{{IEEE} Trans. Wireless
  Commun.}}, vol.~14, no.~10, pp. 5374--5387, Oct. 2015.

\bibitem{duan_d-acb_2016}
S.~Duan, V.~Shah-Mansouri, Z.~Wang, and V.~W.~S. Wong,
  ``\BIBforeignlanguage{en}{{D-ACB}: Adaptive congestion control algorithm for
  bursty {M2M} traffic in {LTE} networks},''
  \emph{\BIBforeignlanguage{en}{{IEEE} Trans. Veh. Technol.}}, vol.~65, no.~12,
  pp. 9847--9861, Dec. 2016.

\bibitem{jin_recursive_2017}
H.~Jin, W.~T. Toor, B.~C. Jung, and J.-B. Seo,
  ``\BIBforeignlanguage{en}{Recursive pseudo-bayesian access class barring for
  {M2M} communications in {LTE} systems},''
  \emph{\BIBforeignlanguage{en}{{IEEE} Trans. Veh. Technol.}}, vol.~66, no.~9,
  pp. 8595--8599, Sep. 2017.

\bibitem{yang_performance_2012}
X.~Yang, A.~Fapojuwo, and E.~Egbogah, ``\BIBforeignlanguage{en}{Performance
  analysis and parameter optimization of random access backoff algorithm in
  {LTE}},'' in \emph{\BIBforeignlanguage{en}{Proc. {IEEE} {Vehicular}
  {Technology} {Conference} ({VTC} {Fall})}}, Quebec City, QC, Canada, Sep.
  2012, pp. 1--5.

\bibitem{chen_dynamic_2020}
J.~Chen, R.-G. Cheng, O.~Agbodike, and Y.-S. Lyu, ``\BIBforeignlanguage{en}{A
  dynamic backoff window scheme for machine-type communications in
  cyber-physical systems},'' \emph{\BIBforeignlanguage{en}{{IEEE} Access}},
  vol.~8, pp. 31\,045--31\,056, Feb. 2020.

\bibitem{althumali_dynamic_2020}
H.~D. Althumali, M.~Othman, N.~K. Noordin, and Z.~M. Hanapi,
  ``\BIBforeignlanguage{en}{Dynamic backoff collision resolution for massive
  {M2M} random access in cellular {IoT} networks},''
  \emph{\BIBforeignlanguage{en}{{IEEE} Access}}, vol.~8, pp.
  201\,345--201\,359, Nov. 2020.

\bibitem{li_dynamic_2015}
W.~Li, Q.~Du, L.~Liu, P.~Ren, Y.~Wang, and L.~Sun,
  ``\BIBforeignlanguage{en}{Dynamic allocation of rach resource for clustered
  {M2M} communications in {LTE} networks},'' in
  \emph{\BIBforeignlanguage{en}{Proc. {International} {Conference} on
  {Identification}, {Information}, and {Knowledge} in the {Internet} of
  {Things} ({IIKI})}}, Beijing, China, Oct. 2015, pp. 140--145.

\bibitem{hwang_dynamic_2015}
H.-Y. Hwang, S.-M. Oh, C.~Lee, J.~H. Kim, and J.~Shin,
  ``\BIBforeignlanguage{en}{Dynamic {RACH} preamble allocation scheme},'' in
  \emph{\BIBforeignlanguage{en}{Proc. {International} {Conference} on
  {Information} and {Communication} {Technology} {Convergence} ({ICTC})}}, Jeju
  Island, South Korea, Oct. 2015, pp. 770--772.

\bibitem{shahin_hybrid_2018}
N.~Shahin, R.~Ali, and Y.-T. Kim, ``\BIBforeignlanguage{en}{Hybrid
  slotted-{CSMA}/{CA}-{TDMA} for efficient massive registration of {IoT}
  devices},'' \emph{\BIBforeignlanguage{en}{{IEEE} Access}}, vol.~6, pp.
  18\,366--18\,382, Mar. 2018.

\bibitem{pratas_code-expanded_2012}
N.~K. Pratas, H.~Thomsen, C.~Stefanovic, and P.~Popovski,
  ``\BIBforeignlanguage{en}{Code-expanded random access for machine-type
  communications},'' in \emph{\BIBforeignlanguage{en}{Proc. {IEEE} {Globecom}
  {Workshops}}}, Anaheim, CA, USA, Dec. 2012, pp. 1681--1686.

\bibitem{kim_efficient_2017}
J.~S. Kim, S.~Lee, and M.~Y. Chung, ``\BIBforeignlanguage{en}{Efficient
  random-access scheme for massive connectivity in {3GPP} low-cost machine-type
  communications},'' \emph{\BIBforeignlanguage{en}{{IEEE} Trans. Veh.
  Technol.}}, vol.~66, no.~7, pp. 6280--6290, Jul. 2017.

\bibitem{Liu2018}
L.~Liu and W.~Yu, ``Massive connectivity with massive {MIMO—Part I}: {Device}
  activity detection and channel estimation,'' \emph{{IEEE} Trans. Signal
  Process.}, vol.~66, no.~11, pp. 2933--2946, Jun. 2018.

\bibitem{Han2020}
H.~Han, Y.~Li, W.~Zhai, and L.~Qian, ``A grant-free random access scheme for
  {M2M} communication in massive {MIMO} systems,'' \emph{{IEEE} Trans. Signal
  Process.}, vol.~7, no.~4, pp. 3602--3613, Apr. 2020.

\bibitem{rusek_scaling_2013}
F.~Rusek, D.~Persson, {Buon Kiong Lau}, E.~G. Larsson, T.~L. Marzetta, and
  F.~Tufvesson, ``\BIBforeignlanguage{en}{Scaling up {MIMO}: Opportunities and
  challenges with very large arrays},'' \emph{\BIBforeignlanguage{en}{{IEEE}
  Signal Process. Mag.}}, vol.~30, no.~1, pp. 40--60, Jan. 2013.

\bibitem{larsson_massive_2014}
E.~G. Larsson, O.~Edfors, F.~Tufvesson, and T.~L. Marzetta, ``Massive {MIMO}
  for next generation wireless systems,'' \emph{{IEEE} Commun. Mag.}, vol.~52,
  no.~2, pp. 186--195, Feb. 2014.

\bibitem{xiong_group-based_2018}
X.~Xiong, L.~Hou, and L.~Zhao, ``\BIBforeignlanguage{en}{A group-based massive
  multiple access scheme in cellular {M2M} networks},''
  \emph{\BIBforeignlanguage{en}{Elsevier Computer Communications}}, vol. 121,
  pp. 44--49, May 2018.

\bibitem{carvalho_random_2017}
E.~d. Carvalho, E.~Bjornson, J.~H. Sorensen, P.~Popovski, and E.~G. Larsson,
  ``Random access protocols for massive {MIMO},'' \emph{{IEEE} Commun. Mag.},
  vol.~55, no.~5, pp. 216--222, May 2017.

\bibitem{sutton_reinforcement_1998}
R.~S. Sutton and A.~G. Barto, \emph{\BIBforeignlanguage{en}{Reinforcement
  Learning: An Introduction (2nd Edition)}}.\hskip 1em plus 0.5em minus
  0.4em\relax Cambridge, MA: MIT Press, 2018.

\bibitem{hasselt_deep_2016}
H.~v. Hasselt, A.~Guez, and D.~Silver, ``Deep reinforcement learning with
  double {Q}-learning,'' in \emph{Proc. AAAI Conference on Artificial
  Intelligence}, Phoenix, AZ, USA, Feb. 2016, pp. 2094--2100.

\bibitem{balanis_antenna_2016}
C.~A. Balanis, \emph{Antenna Theory: Analysis and Design (4th Edition)}.\hskip
  1em plus 0.5em minus 0.4em\relax Hoboken, NJ: Wiley, 2016.

\end{thebibliography}
%
%
%
\end{document}